\documentclass[aps, prd, twocolumn, lengthcheck, superscriptaddress, showpacs, letterpaper, nofootinbib]{revtex4-1}

\usepackage{amsfonts}
\usepackage{amsmath}
\usepackage{amssymb}
\usepackage{xcolor}
\usepackage{graphicx}
\usepackage{subfigure}
\usepackage{slashed}
\usepackage{hyperref}
\usepackage{amsthm}
\usepackage{mathrsfs}
\usepackage{color}
\usepackage{epsfig}
\usepackage{epstopdf}
\usepackage{multirow}
\usepackage{bm}

\def\be{\begin{eqnarray}}
\def\ee{\end{eqnarray}}

\def\lsim{\mathrel{\rlap{\lower3pt\hbox{\hskip1pt$\sim$}}
		\raise1pt\hbox{$<$}}} 
\def\gsim{\mathrel{\rlap{\lower3pt\hbox{\hskip1pt$\sim$}}
		\raise1pt\hbox{$>$}}} 

\allowdisplaybreaks

\begin{document}

\title{Multi-skyrmion states in the Skyrme model with false vaccum potential}

\author{Jun-Shuai Wang}
\affiliation{College of Physics, Jilin University, Changchun, 130012, China}

\affiliation{School of Fundamental Physics and Mathematical Sciences,
	Hangzhou Institute for Advanced Study, UCAS, Hangzhou, 310024, China}

\author{Yong-Liang Ma}
\email{ylma@ucas.ac.cn}
\affiliation{School of Fundamental Physics and Mathematical Sciences,
	Hangzhou Institute for Advanced Study, UCAS, Hangzhou, 310024, China}

\begin{abstract}
We study the multi-skyrmion states using a Skyrme model with false vacuum potential upto baryon number $B=8$ using the product ansatz. It is found that, both the false vacuum potential and true vacuum potential can yield cluster structure of the multi-skyrmion states. The effect of the explicit chiral breaking on the masses and the contour surfaces of the baryon number density of the multi-skyrmion states are analyzed.
\end{abstract}

\maketitle


In the 1960s, T. R. Skyrme put forward a pioneer idea to study nucleon physics using a nonlinear theory of mesons which is named as Skyrme model later~\cite{Skyrme:1961vq,Skyrme:1962vh}. Nowdays, the Skyrme model and its extension have been widely used in nuclear physics and condensed matter physics~\cite{Zahed:1986qz,RhoZahed,Ma:2016gdd,Ma:2019ery,Manton:2022fcb}.

The original Skyrme model is very simple. It only consistes the nonlinear sigma model term and the Skyrme term which is actually a combination of the $L_1$ and $L_2$ terms in the later developed chiral perturbation theory~\cite{Gasser:1983yg}. In literature, the Skyrme model was extended for various purpose by, for exmple, including by the pion mass, the hadron resonances or saturating the Bogomolnyi's bound~\cite{Jackson:1983bi,Adkins:1983hy,Jackson:1985yz,Meissner:1986vu,Marleau:1990nh,Adam:2010ds,Adam:2013wya,Sutcliffe:2011ig,Ma:2012zm,Ma:2012kb,Gudnason:2016tiz}.

It is easy to imagine that the minimal extension of the Skyrme model is to include the pion mass term $ m^2\text{Tr}(1-U) $~\cite{Adkins:1983hy}. In literature, the pion mass effect has been studied for several years~\cite{RhoZahed,Battye:2004rw, Battye:2006tb, Gillard:2015eia,Manton:2022fcb}. People found that, the pion mass term affects the single skyrmion properties and globally, improves the results to the empirical values~\cite{Adkins:1983hy}. In the multi-skyrmion system which the present work focuses on, the pion mass affects the multi-skyrmion spectrum and configurations strongly~\cite{Battye:2004rw,Battye:2006tb,Battye:2006na,Naya:2018kyi}.

Genrally speaking, the $ m^2\text{Tr}(1-U) $ term is not the only one contributing to the pion mass. In principle, a term like $ m^2\text{Tr}(1-U^2) $ also explicitly breaks chiral symmetry. It is found that this term introduces a false vacuum to the Skyrme model~\cite{Dupuis:2018utr,Ferreira:2021ryf} in addition to the vacuum of QCD. So that it is interesting to study the skyrmion properties with respect to the  this false vacuum which not only is interested in hadron physics but also may help us to undestsnd the cosmological phase transition process~\cite{Frampton:1976kf,Guth:1980zm,Guth:1982pn}.

When this false vacuum is added, the Skyrme model admits a false meta-stable skyrmion configuration~\cite{Kumar:2010mv} which can decay due to the tunneling effect~\cite{Lee:2013ega} and consequently causes the decay of the false vacuum~\cite{Dupuis:2018utr}. In addition, it found that with the appropriate choice of the false vacuum potential, the experimental values of radii and binding energies for a very wide range of the mass numbers of nuclei can be calculated with a good precision~\cite{Ferreira:2021ryf}. In Ref.~\cite{Livramento:2022zly}, people studied the multi-skyrmion properties using rational map ansatz.


In this paper, we shall study the multi-skyrmion configurations using the product ansatz for the purpose to see the false vacuum effect on cluster structure of nuclei. We start from the original Skyrme model~\cite{Skyrme:1961vq,Skyrme:1962vh},
\be
\mathcal{L}_{\rm Skyrme} = \frac{f_{\pi}^2}{16} \text{Tr} \left[ \partial_{\mu} U^{\dagger} \partial^{\mu} U \right] + 
	\frac{1}{32 e^2} \text{Tr} \left[ U^{\dagger} \partial_{\mu} U, U^{\dagger} \partial_{\nu} U \right]^2,
\nonumber\\
\label{eq:LSkyrme}
\ee
where $f_\pi=184~$MeV, $e$ is the Skyrme parameter and $U(\mathbf{x})=\exp(2i\pi/f_\pi)=\exp(2i\pi^a\tau^a/f_\pi)$ with $\tau^a$ being the Pauli matrices.

Since the chiral field $U(x)$ is unitary,  for any fixed time, say, $t_0$, the static configuration $U(\mathbf{x}, t_0)$ defines a map from the
manifold $R^3$ to the manifold $S^3$ in the isospin space, that is, 
\be
U(\mathbf{x}, t_0) : R^3 \to S^3.
\label{eq:map}
\ee
At low energy limit, QCD goes to the vacuum, i.e., $U(\mathbf{x}\to \infty, t_0) = 1$, all the points at $\mathbf{x} \to \infty$ are mapped onto the north pole of $S^3$. Therefore, maps \eqref{eq:map} constitute the third homotopy group $\pi_3(S^3) =Z$  where the integer $Z$ accounts for the times that $S^3$ is covered by the mapping $U(\mathbf{x}, t_0)$, i.e., winding numbers. Since the winding number is conserved when the time coordinate is changed and the $N_c$ scaling of the energy of model \eqref{eq:LSkyrme} is the same as baryons in the constituent quark model, the solution of~\eqref{eq:LSkyrme}---soliton---can be regarded as a baryon~\cite{Witten:1979kh}. 

In the following, for convenience, we use the energy and length scales $ { f_{\pi} }/{4e} $ and $ { 2 }/{ef_{\pi}} $. Using these new scales, one can rewrite the Skyrme model Lagrangian~\eqref{eq:LSkyrme} as
\begin{equation}
	\mathcal{L}_{\rm Skyrme} = \frac{1}{2} \text{Tr} \left[ \partial_{\mu} U^{\dagger} \partial^{\mu} U \right] + 
	\frac{1}{16} \text{Tr} \left[ U^{\dagger} \partial_{\mu} U, U^{\dagger} \partial_{\nu} U \right]^2.
	\label{eq:SkyrmeNS}
\end{equation}
Then, in the new scales, the extended Skyrme model including the false vacuum we will use has the form
\be
\mathcal{L}_{\rm ESkyr} = \mathcal{L}_{\rm Skyrme} + \mathcal{L}_{mass},
\label{eq:ESkyrL}
\ee
with
\be
\mathcal{L}_{mass} ={} -\frac{1}{4} \left( m_1^2 \text{Tr}\left[1-U\right] + m_2^2 \text{Tr}\left[1-U^2\right] \right).
\label{eq:ESkyrMass}
\ee

To study the effect of the false vacuum potential on the multi-skyrmions, in particular the
effect of the cluster structure, we decompose the static field $U(x)$ as
\be
U( \mathbf{x} ) = \sigma + i \bm{\tau} \cdot \bm{\pi},
\label{eq:ParamU}
\ee
where $ \bm{\tau}=( \tau_1, \tau_2, \tau_3 ) $ are the Pauli matrices, $ \bm{\pi} = ( \pi_1, \pi_2, \pi_3 ) $
are the Goldstone bosons after chiral symmetry breaking. The chiral fields $\sigma$ and $\pi_i$ satisfy the constraint
\be
\sigma^2 + \bm{\pi} \cdot \bm{\pi} = 1.
\label{eq:SPConstraint}
\ee

\begin{figure}[tbh]
\centering
\includegraphics[width=8cm]{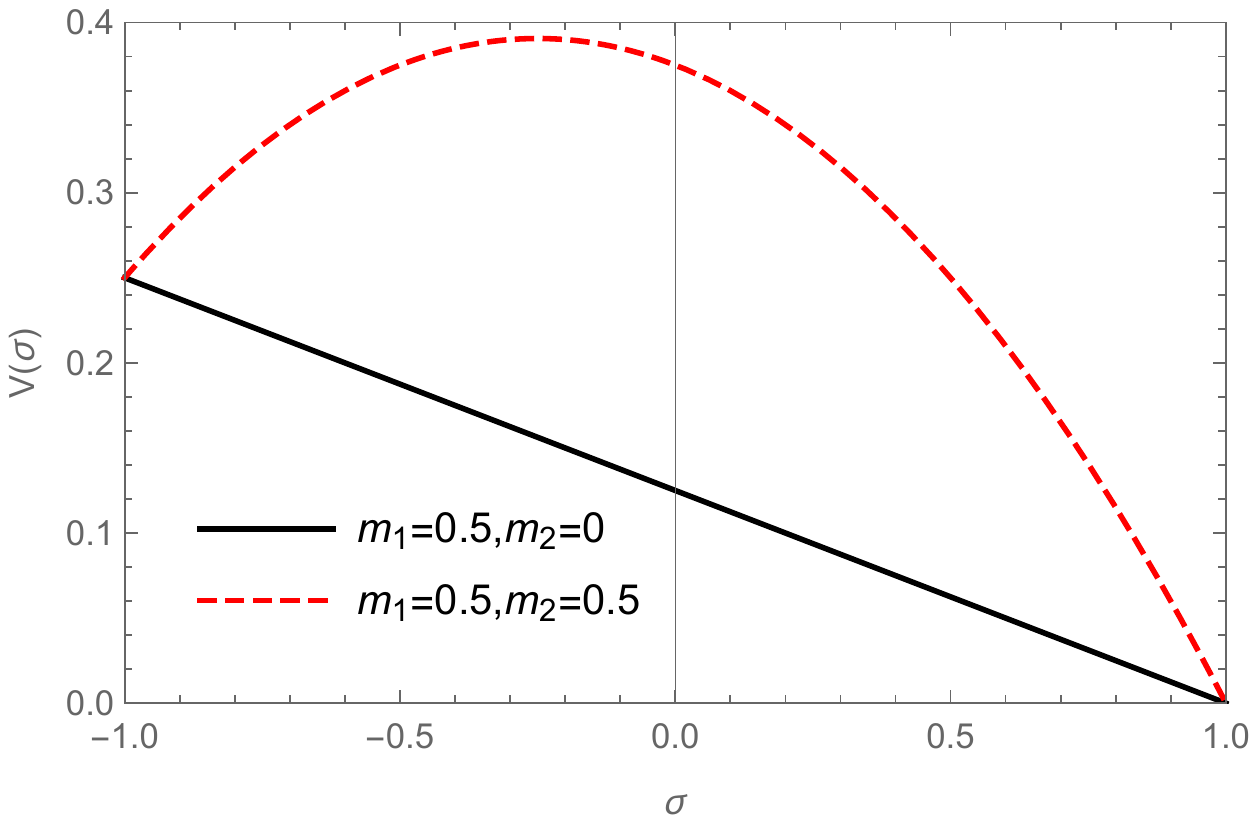}
\caption{Cartoon of false vacuum potential.}
\label{fig:FalsePotential}
\end{figure}

With parametrization~\eqref{eq:ParamU}, the false vacuum potential term can be approximated as, 
\be
V = \frac{1}{2} m_1^2 (1-\sigma) + m_2^2 (1-\sigma^2).
\label{eq:potential}
\ee
It follows that, when $ m_1 ^2 > 4m_2 ^2 $, e.g., $m_1=0.5, m_2=0$, there is only one global minimum located at $ U=\sigma=1$. However, when $ m_1 ^2 < 4m_2 ^2 $, e.g., $m_1=0.5, m_2=0.5$, in addition to the global minimum, there is a local minimum at $ U= \sigma = {} - 1$. This is illustrated in Fig.~\ref{fig:FalsePotential}. It should be noted that, due to the quantum tunneling effect, the local minimum is metastable.

From Lagrangian~\eqref{eq:ESkyrL} and using the parameterization~\eqref{eq:ParamU}, we obtain the Hamiltonian of the system as
\be
	H & = & \int d^3x
	\Big \{ \partial_{i} \sigma \partial_i \sigma + \partial_{i} \vec{\pi} \cdot \partial_i \vec{\pi} \nonumber\\
	& &\qquad {} - \frac{1}{2} \left( \partial_{i} \sigma \partial_j \sigma + 
	\partial_{i} \vec{\pi} \cdot \partial_j \vec{\pi} \right) \left( \partial_{i} \sigma \partial_j \sigma + 
	\partial_{i} \vec{\pi} \cdot \partial_j \vec{\pi} \right) \nonumber\\
	& &\qquad {} +\frac{1}{2} \left( \partial_{i} \sigma \partial_i \sigma + 
	\partial_{i} \vec{\pi} \cdot \partial_i \vec{\pi} \right) \left( \partial_{j} \sigma \partial_j \sigma + 
	\partial_{j} \vec{\pi} \cdot \partial_j \vec{\pi} \right) \nonumber\\
	& &\qquad {} + V_{eff} \Big \}
\label{eq:energy}
\ee
where for convenience we introduce the effective potential $ V_{eff} $ as
\be
V_{eff} & \equiv &
		\begin{cases}
			V, \quad & U(\infty)= + 1~ ( \text{skyrmions} )\\
			V-m_1^2, & U(\infty)= - 1~ ( \text{false skyrmions} )
		\end{cases}
\ee
which satisfies $V_{eff}\to 0$ as $\mathbf{x} \to \infty$.


We next study the effect of the flase vaccum potential on the multi-skyrmion configurations using the product ansatz. The basic idea of product ansatz can be summarized as follows: Given two $SU(2)$ symmetric fields---the skyrmion solutions, the one with baryon number $B_1$ called $ U_1 (\mathbf{x}_1) $ and the other one with baryon number $ B_2 $ called $ U_2 (\mathbf{x}_2) $, one can construct a new skyrmion configuration $U(\mathbf{x})$ with baryon number $ B=B_1 + B_2 $ by multiplying them together. To find the lowest energy configuration of the new skyrmion $U(\mathbf{x})$, the two original skyrmions are placed with a relative rotation in the isospin space
\be
 U(\mathbf{x}) = U_1 (\mathbf{x}_1) C(\bm{\alpha}) U_2 (\mathbf{x}_2) C ^\dagger(\bm{\alpha}) \equiv U_1 (\mathbf{x}_1) U_2^\prime (\mathbf{x}_2),
\ee
where $ C(\bm{\alpha}) =\text{exp} ( i \bm{\tau} \cdot \bm{\alpha} /2 ) = \text{exp} ( i \bm{\tau} \cdot \hat{ \bm{\alpha} } \alpha /2 ) $, representing a rotation of $\alpha$ angles around $ \hat{ \bm{\alpha} } $-axis in the isospin space. The direction of the $\hat{\bm{\alpha}}$-axis and the magnitude of the roration angle $\alpha$ are determined by minimizing the energy of the system which a physical system should satisfy.By using the parametrization~\eqref{eq:ParamU}, the new $U^\prime$ field after rotation is expressed as
\be
U^\prime(\mathbf{x}) & = & \sigma' + i \bm{\tau} \cdot \bm{\pi}'( \mathbf{x} ) 
\ee
with
\be
\sigma^\prime & = & \sigma    \nonumber \\
\bm{\pi}^\prime & = & ( \bm{\pi} \times \hat{ \bm{\alpha} } ) \sin( \alpha ) + \bm{\pi} \cos( \alpha ) + \hat{ \bm{\alpha} } \bm{\pi} \cdot \hat{ \bm{\alpha} } ( 1-\cos( \alpha ) ).
\nonumber\\
\ee
The multi-skyrmion states are obatined by varying the rotation angle and the distance between the initial states.


In the numerical calculation, in order to find the energy minima of the multi-skyrmions system, the finite element method~\cite{buelerPETScPartialDifferential2020} is used to solve the three-dimensional partial differential equations generated from the variation of Eq.~\eqref{eq:energy}, with an additional Lagrange multiplier $\lambda$ to constraint~\eqref{eq:SPConstraint}.

\begin{figure}[tbh]
	\centering
	\includegraphics[width=8.0cm]{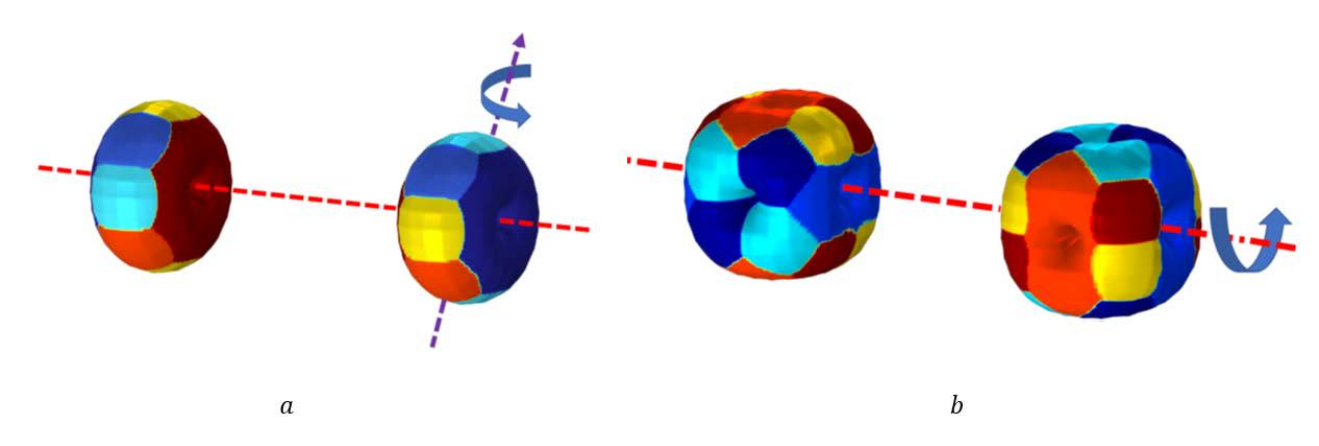}
	\caption{Ralative rotations of the $B=2$ clusters (left panel) and $B=4$ clusters (right panel).}
	\label{fig:rotation}
\end{figure}

The initial values of the multi-skyrmions solutions are constructed by using the product ansatz~\cite{Battye:1996nt, Salmi:2015wvi}. Explicitly, the initial value of the  baryon number $B=2$ state is constructed using the solutions of the two $B=1$ states---clusters---which are far away from each other. The relative rotation of the two clusters are found to have angle $\pi$ along the axes perpendicular to the line connecting the two clusters. Similarly, for the baryon number $B=4$ state, its initial value is obtained by producting the previous solutions of the $B=2$ skyrmions. However, since the $B=2$ clusters are  not ideal spheres, the relative rotations between them are complicated. It is found that, to have the strongest attraction, the two clusters should rotate angle $\pi$ along the axis perpendicular to the line crossing the holes of the two donuts, as shown in Fig.~\ref{fig:rotation}a. The initial value of the baryon number $B=8$ state is constructed from two skyrmions of $B=4$ with relative rotation $\pi/2$, as shown in Fig.~\ref{fig:rotation}b.

It should be noted that the initial value of the multi-skyrmion state can also be obtained by using the vlues of each cluster given by the rational map ansatz~\cite{Houghton:1997kg}. The relative rotation between the two clusters are found the same as that in the product ansatz. The final results yielded from these two approach agree to each other.

The use of product ansatz only constructs the initial value of the multi-skyrmion state with the two clusters far away from each other. A possible way to obtain the bound state in the attractive channel is to approach the two clusters by adding a pseudo-time component to the static configuation, which is the so called relaxation method~\cite{Salmi:2015wvi, Battye:1996nt}. Here, we use a simpler method to reduce the distance by squeezing the original configuation. The method can be understood as follows using a one-dimensional ansatz: Let the size of the final multi-skyrmion state with profile $f(x)$ to be $L$, the size of each cluster is $L/2$ and the center of the two clusters is origin. In the next iteration, the size of the one dimensional soliton becomes $L - dl$, and the size of each
cluster is $L/2-dl/2$, and the profile function in the next iteration can be defined as, 
\be
f^\prime(x) & = & 
        \begin{cases}
            f( x - \frac{dl}{2} ), & -\frac{L}{2}+\frac{dl}{2} \leqslant x < -\frac{L}{4} + \frac{dl}{2} \\
            f( \frac{L}{L-2dl} x ), & |x| \leqslant \frac{L}{4} - \frac{1}{2} dl \\
            f( x + \frac{dl}{2} ), & \frac{L}{4}-\frac{dl}{2} < x \leqslant \frac{L}{2}-\frac{dl}{2}
        \end{cases}.
\ee
In this way, the size of the solitons decreases by $dl$ for each iteration. The final profile is obtained until the energy minima is accessed. In this work the two clusters of skyrmions are aligned in the $x$-direction, and the parameters $dl=0.02$ and $dl=0.05$ are used for obtainning the minima.

To show the effect of the mass parameters on the multi-skyrmion configurations, we use four typical combinations,
\begin{itemize}
	\item $m_1 = m_2 = 0.5$: Skyrme model with false vacuum.
	\item $m_1 =0, m_2 = 0.5$: Skyrme model with false vacuum.
	\item $m_1 =0.5, m_2 = 0$: Skyrme model with massive pions.
	\item $m_1 = m_2 = 0$: Skyrme model in chiral limit.
\end{itemize}

\begin{figure}[htbp] \centering
	\includegraphics[width=1.0\linewidth]{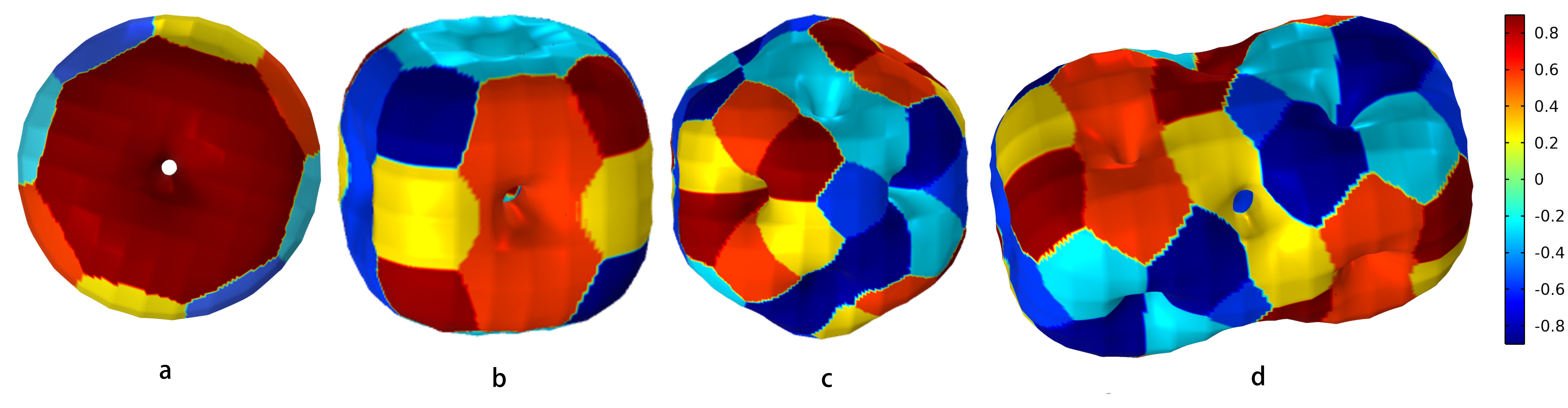}
	\caption{Baryon number density $\mathcal{B}^0=0.02$ contour suface for false vacuum $m_1=m_2=0.5$. Colors 
		encoding the largest magnitude and the sign of the three constiuent pion fields. a: false skyrmions 
		with $B=2$. b: false skyrmions with $B=4$. c: truncated octahedron false skyrmions with $B=8$. 
		d: $D_{4h}$ false skyrmions with $B=8$.
	} 
	\label{fig:false248}
\end{figure}

Figure~\ref{fig:false248} shows the contour surface of baryon number density with $m_1 = m_2 = 0.5$ for multi-skyrmion states with baryon numbers $B=2,4,8$. For baryon numbers $B=2,4$, one can see that the symmetries of false skyrmions calculated in this work  (see 
Fig.~\ref{fig:false248}a and Fig.~\ref{fig:false248}b) are the same as that of the true skyrmions~\cite{Houghton:1997kg}. However, for the multi-skyrmion state with baryon number $B=8$, we obtain two configurations Fig.~\ref{fig:false248}c using $dl=0.02$ and Fig.~\ref{fig:false248}d using $dl=0.05$ corresponding, respetively, to the truncated octahedron solution and $D_{4h}$ solution. The latter can be regarded as a bound state consists of two cubic skyrmions with $B=4$. In our calculation, the $D_{4h}$ solution is a local minima, while the truncated octahedron solution is a global minima.

\begin{figure}[htp] \centering
    \includegraphics[width=0.8\linewidth]{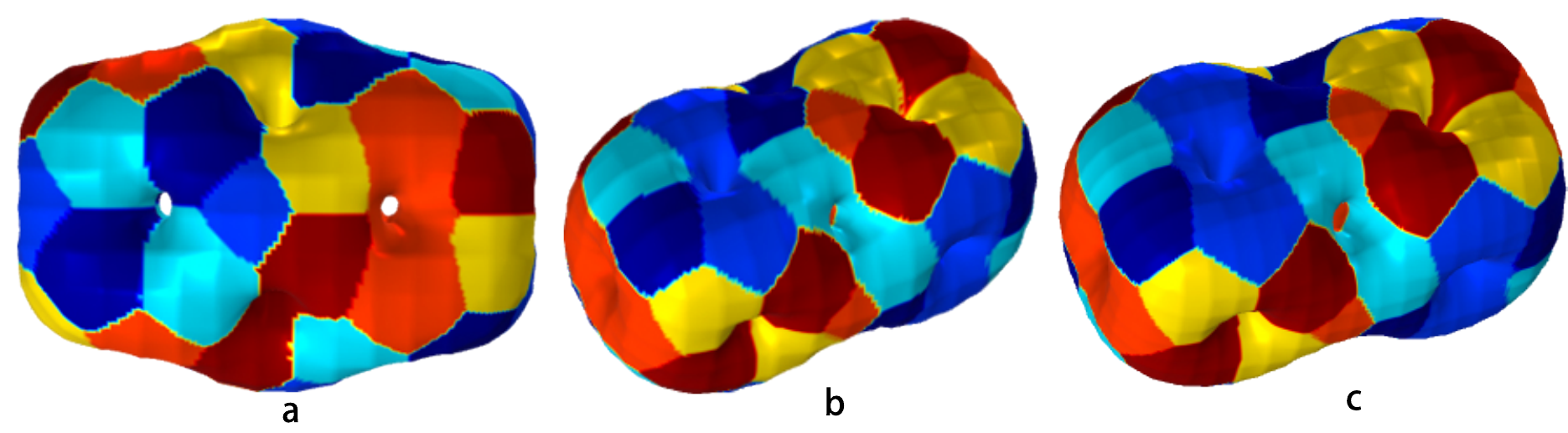}
    \caption{Baryon number density $\mathcal{B}^0=0.02$ contour suface for $B=8$ state. a: false skyrmions with $m_1=0, m_2=0.5$; b: skyrmions with $m_1=0.5, m_2=0$; c: skyrmions with $m_1=0, m_2=0$.
    } 
    \label{figure_b8_else}
\end{figure}

Although the two configurations of the $B=8$ state with $m_1=m_2=0.5$ depend on the choice of $dl$, this penomena does not happen for other sets of mass parameters, at least for the two values of $dl$ choosen. This is confirmed by the calculation using the other three sets of mass parameters, as shown in Fig.~\ref{figure_b8_else}. It is also worth noting that for the mass parameters $m_1=0, m_2=0.5$, as shown in Fig.~\ref{figure_b8_else}a, the two clusters are closer than that for other two sets and have a more prominent contact. Also in the case of the mass parameters $m_1=0, m_2=0$, i.e., in the standard skyrme model, the skyrmion solution with baryon number $B=8$ has two clusters with $B=4$ (Fig.~\ref{figure_b8_else}c). This cluster structure of the $B=8$ multi-skyrmion states may arise from the product ansatz used here. As studied in Ref.~\cite{Battye:2006na}, the truncated octahedron structure may have lower energy by using the rational map approach.

We present in Table.~\ref{tab:energy} the energy of the multi-skyrmion system with different choice of the mass parameters. One can clearly see that the inclusion of the explicit chiral symmetry breaking effect increases both the masses of the multi-skyrmion states and decreases the binding energies. It should be noted that, with all the parameter choices, the bound states exist.

\begin{table}[!h]
	\centering
	\caption{Energy for multi-skyrmion states (scale $12\pi^2\cdot f_\pi/4e$).}
	 \label{tab:energy}
	\begin{tabular}{c|c|c|c|c}
		\hline
		\hline
		\multirow{2}{*}{~$B$~}~& \multicolumn{1}{|c}{~$m_1=0.5$}~&\multicolumn{1}{|c}{ $m_1=0$}~&\multicolumn{1}{|c}{ $m_1=0.5$}~&\multicolumn{1}{|c}{ $m_1=0$}\cr 
		&$m_2=0.5$~&~$m_2=0.5$~&~$m_2=0$~&~$m_2=0$~
		\cr
		\hline
		~$1$~ &{}1.27     &1.29&1.25&1.23\cr\hline
		~$2$~ &{}2.48     &2.51&2.46&2.42\cr\hline
		~$4$~ &{}4.56     &4.62&4.54&4.47\cr\hline
		~$8$~ &{}8.91/9.00&8.95&9.20&8.86\cr
		\hline
		\hline
	\end{tabular}
\end{table}


In summary, in this work, we add a false vacuum potential term to the skyrme model and focus on the masses and baryon number density distributions of the multi-skyrmion states upto baryon number $B=8$. We found that, using the product ansatz, both the false vacuum potential and the true vacuum potential can yield cluster structures of multi-skyrmion states.  With certain mass parameters, the conclusion may different from that by using the rational map ansatz. This issue should be clarified in detail later.

\acknowledgments	

We would like to thank Y. Tian and H. B. Zhang for their valuable discussions. The work of Y.~L. M. was supported in part by National Science Foundation of China (NSFC) under Grant No. 11875147 and No. 12147103.

\bibliography{Reference}

\end{document}